\documentclass[aps,prb,reprint,superscriptaddress]{revtex4-1}

\usepackage[utf8]{inputenc}

\usepackage{graphicx}
\usepackage{longtable}
\usepackage{amsmath}
\usepackage{subfigure}

\usepackage{color}

\begin{document}


\title{Thermally activated conductance in arrays of small Josephson junctions}


\author{J. Zimmer}
\affiliation{Physikalisches Institut, Karlsruhe Institute of Technology}

\author{N. Vogt}
\affiliation{Institut f{\"u}r Theorie der Kondensierten Materie, Karlsruhe Institute of Technology}

\author{A. Fiebig}
\affiliation{Physikalisches Institut, Karlsruhe Institute of Technology}

\author{S. V. Syzranov}
\affiliation{Institut f{\"u}r Theorie der Kondensierten Materie, Karlsruhe Institute of Technology}

\author{A. Lukashenko}
\affiliation{Physikalisches Institut, Karlsruhe Institute of Technology}

\author{R. Sch{\"a}fer}
\affiliation{Institut f{\"u}r Festk{\"o}rperphysik, Karlsruhe Institute of Technology}

\author{H. Rotzinger}
\email[]{hannes.rotzinger@kit.edu}
\affiliation{Physikalisches Institut, Karlsruhe Institute of Technology}

\author{A. Shnirman}
\affiliation{Institut f{\"u}r Theorie der Kondensierten Materie, Karlsruhe Institute of Technology}

\author{M. Marthaler}
\affiliation{Institut f{\"u}r Theoretische Festk{\"o}rperphysik, Karlsruhe Institute of Technology}

\author{A. V. Ustinov}
\affiliation{Physikalisches Institut, Karlsruhe Institute of Technology}
\affiliation{Russian Quantum Center, 100 Novaya St., Skolkovo, Moscow region, 143025, Russia}


\date{\today}

\begin{abstract}
We present measurements of the temperature-dependent conductance for series arrays of small-capacitance SQUIDs.  At low bias voltages, the arrays exhibit a strong Coulomb blockade, which we study in detail as a function of temperature and Josephson energy $E_J$. We find that the zero-bias conductance is well described by thermally activated charge transport with the activation energy on the order of $\Lambda E_C$, where $\Lambda$ is the charge screening length in the array and $E_C$ is the charging energy of a single SQUID.
\end{abstract}


\maketitle

\section{Introduction}

Chains of ultra-small superconducting islands weakly connected via Josephson junctions are primarily interesting due to their charge transport properties, which are resembling the superconductor to insulator transition identified in thin disordered superconducting films \cite{Goldman_2010}.
Experimental studies\cite{Haviland1996Co,Chow_1998,Gershenson,Pop} of transport in Josephson chains have been also motivated by the possibility of metrological applications as current standards\cite{Piquemal_2004}, theoretical predictions of solitonic phenomena\cite{Hermon1996,Gurarie}, and predictions of topologically protected states in such systems\cite{Doucot:newstate}.

In this paper, we report a quantitative study of the thermally activated conductance for two arrays with different number of islands $N=59$ and $N=255$. The islands in the array are coupled by pairs of Josephson junctions connected in parallel and thus forming superconducting quantum interference devices (SQUIDs). Two important physical properties of the array are the charge screening length $\Lambda= \sqrt{C/C_0} $ measured in the number of islands, and the junction charging energy  $E_C = 2e^2/C $. Here, $C$ and $C_0$ are the electrostatic capacitance between neighboring islands and the island capacitance to the ground, respectively.  For the studied arrays, the values $\Lambda\approx 10$ and $E_C\approx 44\;\mu\mathrm{eV}$ have been estimated from scanning electron micrographs and literature values for the specific junction capacitances\cite{Watanabe2001Co}. The typical energy needed to push a Cooper pair into the array is of the order of $\Lambda E_C$.

\begin{figure}[!ht]
\begin{center}
\includegraphics[width=.4\textwidth]{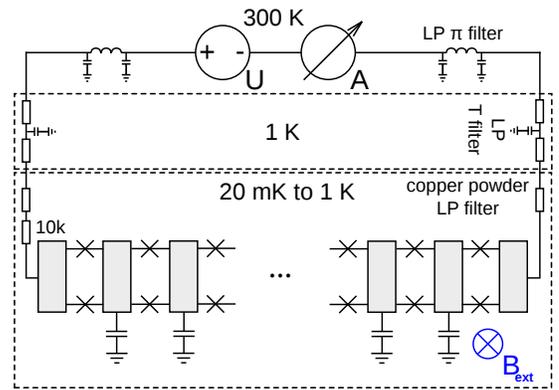}
\end{center}
\caption{\label{fig_setup_a}
Sketch of experimental setup. A chain of small-capacitance SQUIDs is measured using a voltage-bias scheme. The sample leads are filtered at several stages of the dilution refrigerator. 
}
\end{figure}

The Josephson coupling energy of two Josephson junctions connected in parallel is at maximum $2E_\mathrm J= \Phi_0 2I_C/2\pi$. The coupling between the islands is tuned by applying an external magnetic field, see Fig.~\ref{fig_setup_a}. Here $\Phi_0$ is the magnetic flux quantum and $I_C$ is the critical current of a single junction. The zero-field Josephson energy $2E_{J}$ is estimated by measuring the tunneling resistance of the arrays at high bias voltages and employing the Ambegaokar-Baratoff formula\cite{Ambegaokar1963Er}. We thus obtain $2E_{J} = 140\pm20\;\mu\mathrm{eV}$. Having the screening length $\Lambda$  smaller than the length of the array makes its interior well protected from the environment by the electrostatic screening of the outer islands. To gain information on the low-bias conductance properties of the whole array, we study charge transport in our samples at small bias voltages, for different temperatures and varied $E_\mathrm J$.

Measurements of the current-voltage characteristics at base temperature $T\approx 20$ mK display a Coulomb blockade of the current up to voltages on the order of a millivolt, depending on the ratio between the charging energy $E_C$ and the magnetic field-dependent Josephson energy $E_J $, in agreement with previous reports\cite{Haviland1996Co}. At temperatures exceeding $200$ mK, the Coulomb blockade is gradually lifted and a finite differential conductance at zero bias develops. In this work, we mainly focus on this transition region. 

\section{Theoretical Description}\label{sec_TheoDesc}

The effective Hamiltonian of Cooper pairs in a chain of Josephson junctions reads\cite{Efetov:fundamental}
\begin{eqnarray}\label{Hamiltonian}
\hat{\mathcal{H}}=\frac{1}{2}\sum_{r,r^\prime}B_{rr^\prime}(\hat{n}_{r}-n_r^x)(\hat{n}_{r^\prime}-n_{r^\prime}^x)-\nonumber\\
-\sum_{r}E_J\cos\left(\phi_{r}-\phi_{r+1}\right),
\end{eqnarray}
where indices $r$ and $r^\prime$ number the superconducting islands between the junctions,
$\phi _{r}$ and $\hat{n}_{r}=-i\partial /\partial
\phi_{r}$ are respectively the phase of the superconducting order
parameter and the operator of the number of Cooper pairs in
island $r$; $B_{rr^\prime}$ is the inverse capacitance matrix of the chain in units of $(2e)^2$;
$B_{rr^\prime}\approx \Lambda E_C\,e^{-|r-r^\prime|/\Lambda}$ for $\Lambda \gg1$.
Random quantities $n_r^x$ account for the stray charges in the junctions and in the
dielectric environment of the chain.

Sub-micron sizes of the superconducting islands, with which we deal in this work,
correspond to large characteristic numbers of stray charges per island\cite{Feigel:review,Serret,Maisi2009Pa},
$|n_r^x|\gg1$.
As we show below, at so strong charge disorder only the fractional parts of $n_r^x$ are important.
Because the offset potential fluctuates from site to site, at small Josephson couplings each Cooper-pair
excitation is localized almost entirely on a single island, and no transport is possible at $T=0$.
At finite temperature localized Cooper pairs can hop inelastically from site to site
by absorbing and emitting neutral excitations, e.g., phonons or Cooper-pair dipoles.
In this section, we consider the zero-bias-voltage conductance of a Josephson chain in the experimentally relevant regime 
of strong disorder. 

It is convenient to count the energies of all Cooper-pair excitations from
the energy of the ``ground charging state''\cite{Syzranov:interf}, the configuration of integers $n_r=n_r^0$,
which minimize the charging energy of the chain, the first term in the right-hand side of Eq.~(\ref{Hamiltonian}).
The simplest excitations on each site correspond to adding an extra Cooper pair to
the ``ground charging state'' (boson) or to removing a Cooper pair (antiboson). 
The energies 
$E_r^+=\frac{1}{2}B_{rr}+\sum_{r^\prime}B_{rr^\prime}(n^0_{r^\prime}-n_{r^\prime}^x)$
and $E_r^-=\frac{1}{2}B_{rr}-\sum_{r^\prime}B_{rr^\prime}(n^0_{r^\prime}-n_{r^\prime}^x)$
of the bosons and the antibosons are of the order of $\Lambda E_C$ on most sites,
but can also have arbitrarily small values due to the presence of disorder.
The (anti)boson tunneling element between neighboring sites
is $\langle r|\hat{\cal H}|r+1\rangle=-E_J/2$.

The characteristic charging energies significantly exceed the temperature,
leading to Boltzmann statistics of the excitations. 
Because only the lowest-energy excitations on each site contribute to conduction,
it is sufficient to assume that each island can host no more than one (anti)boson.
 Under these conditions, the hopping
of the localized bosons and antibosons in the system under consideration
is equivalent to the hopping of electrons and holes in a disordered semiconductor,
which allows us to apply straightforwardly the well-known results
for hopping transport at strong disorder\cite{Shklovskii:book}.

The conduction can be described by mapping the chain of islands
onto the so-called Abrahams-Miller network; to each pair of islands $r$ and $r^\prime$
one has to assign a resistance, which exponentially depends on the distance 
between the sites and the excitation energies,
\begin{equation}
	R_{rr^\prime}^\pm\propto
	\exp\left[2|r-r^\prime|/\xi+\max(E_r^\pm,E_{r^\prime}^\pm)/T\right],
	\label{AMresistor}
\end{equation}
where $\xi$ is the effective localization length
of a single charge excitation; ``+'' and ``-'' refer to the contributions of the bosons and the antibosons
to the conductivity. 

The conductivity then depends exponentially on the temperature\cite{Gantmakher:book,Shklovskii:book}, 
\begin{equation}
	\sigma\propto\exp[-(T_0/T)^\beta],
	\label{sigmagen}
\end{equation}
where $\beta=1$ and $T_0$ is of the order of $\Lambda E_C$
if transport is dominated by the nearest-neighbor (NN) inelastic hops, and
$\beta<1$ if next-to-nearest-neighbor or longer hops are important
[the variable-range hopping (VRH) regime]\cite{Gantmakher:book,Shklovskii:book}.
The behavior of the conductivity crosses over from the NN to the VRH with lowering the temperature
or increasing the Josephson coupling $E_J$. The conductance in the former regime is discussed in
more detail in the Appendix.

These arguments\cite{Shklovskii:book} hold for a bath of
arbitrary nature, so long as it is characterized by Boltzmann statistics, and for an arbitrary
form of the coupling between the bath and the Cooper pairs, which will affect only the
preexponential functions in Eqs.~(\ref{AMresistor}) and (\ref{sigmagen}).
Such phenomenology is robust
enough to predict the exponent $\beta$, but does not allow one to obtain the preexponential
function in Eq.~(\ref{sigmagen}) without making extra assumptions about the nature of the bath and
its microscopic details.


Assuming the bath couples to Cooper pairs locally on each island, the conductance of each junction 
(between the neighboring islands) is proportional to $E_J^2$. Thus, in the regime of NN hopping the total conductance 
will scale as $E_J^2$ and will show the activational behavior ($\beta =1$). Based on the experimental data presented in Section~IV below, we assume our arrays are close to this regime. Thus, we will fit   
the temperature dependence of the conductance as
\begin{equation}\label{eq_G_T}
	G(T) \propto T^{-\alpha}\exp(-T_0/T),
\end{equation}
where the exponent $\alpha$ depends on the
nature of the bath, its spectrum, the form of the boson-bath coupling, etc., and is usually of order unity.

\section{Experimental details \label{setup}}

The experimentally studied arrays consist of Al/AlO$_x$/Al SIS Josephson junctions fabricated using conventional e-beam lithography and the shadow evaporation technique~\cite{Dolan77Of}.
Both investigated arrays were prepared on the same chip with nominally the same fabrication parameters, differing only in the number of SQUIDs, which are N=59 and N=255. The estimated length $N/\Lambda$ of the arrays is thus equal to 5.9 and 25.5, correspondingly.
The charging energy is estimated from the junction area $A = 0.018\;\mu\mathrm{m}^2$, which is measured from scanning electron micrographs. A specific capacitance of $c_s = 200\;\mathrm{fF}/\mu\mathrm{m}^2$ is assumed, which according to Ref.~\citenum{Watanabe2001Co} is much better suited for small junctions than the value of $45\;\mathrm{fF}/\mu\mathrm{m}^2$ commonly used for larger AlO$_x$ based SIS Josephson junctions. 
The Cooper pair charging energy $E_C = 2e^2/C$ is thus estimated to be $44\;\mu\mathrm{eV}$. At temperatures well below 100 mK, the arrays show hysteretic switching out of the Coulomb blockade of current to a finite conductance branch at a critical voltage $V_c$. $V_c$ was measured for the array with N=255 islands and a magnetic flux of $\Phi_0/2$ to be about 1.8~mV; for N=59, the critical voltage is 300 $\mu$V.

 All measurements were performed in a dilution refrigerator with a base temperature of 20 mK. The sample bias was supplied via DC lines with low-pass filters at several stages of the cryostat (see Fig.~\ref{fig_setup_a}). Additional low-pass copper powder filters with a cutoff frequency of about 1~MHz\cite{Lukashenko2008Im} have been located at base temperature close to the sample to suppress residual high-frequency noise. The current sense resolution of the setup was limited by the noise of the room temperature amplifier to $I_{\mathrm{RMS}}\approx 200$~fA.
For each temperature step in the range between 200 and 900 mK the cryostat was stabilized to $\Delta T_{\mathrm{RMS}}/T < 0.1\%$. We measured the current-voltage ($I-V$) characteristics in a voltage-bias scheme (see Fig.~\ref{fig_setup_a}), utilizing least square fits to obtain the small bias conductance (see Fig.~\ref{IV}(a), top inset).

\section{Results}

\begin{figure}[t!]
\begin{center}
\includegraphics[width=0.45\textwidth]{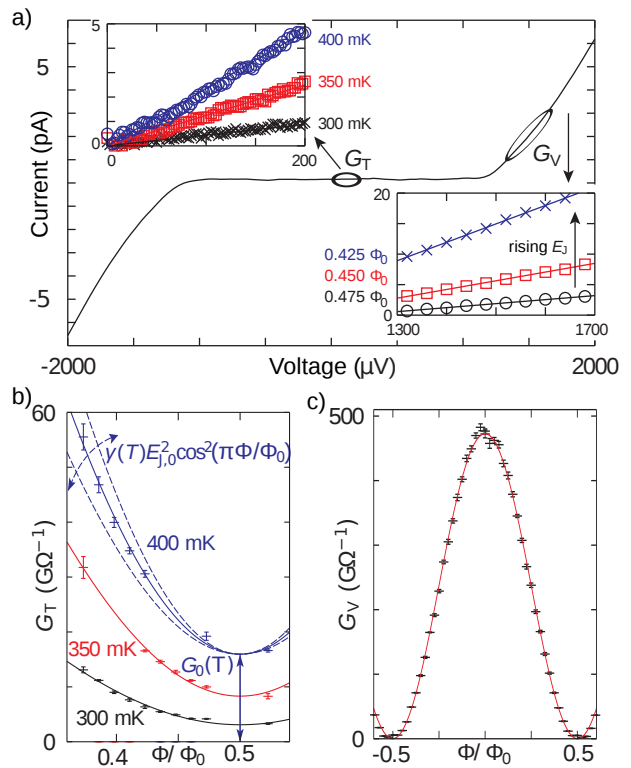}
\caption{\label{IV}
a) Current-voltage characteristic of the 255-islands array at a temperature well below the thermal activation of transport and at an external flux of $\Phi_0/2$ per SQUID loop. Within the setup's resolution, no current can be detected below 1 mV bias voltage due to the Coulomb blockade.
Left inset: with increasing temperature, a finite conductance ($G_T$) develops close to zero voltage bias. Right inset: above a threshold voltage (1.2 mV for N=255 near full suppression of $E_J$), a constant differential conductance ($G_V$) can be observed.
b) Thermally activated conductance $G_T$ in the small-voltage bias range for the array with N=255 islands. The dependence on the external flux has been fitted according to Eq.~(\ref{eq_separation1}) and Eq.~(\ref{eq_separation}). The value of $G_0$ is determined by the offset of the conductance at $\Phi_0/2$, and $\gamma$ captures the magnitude of the $E_J^2$-dependent contribution.
c) Data points showing the differential conductance $G_V$ at temperature 20 mK as a function of flux, in the case of N=255. The solid line is a fit according to $G_V\propto\cos^2(\pi\Phi/\Phi_0)$.
}
\end{center}
\end{figure}

Examining the current-voltage characteristics of both arrays, we find a nearly constant differential conductance $G_V$ in the bias voltage range of several mV indicated in Fig.~\ref{IV}(a), in agreement with previous observations~\cite{Roland_unpublished}. We use the voltage range between 1.3 mV and 1.7 mV to extract the values for $G_V$ as a function of magnetic flux $\Phi$, shown in Fig.~\ref{IV}(c). 

The external flux $\Phi$ applied to SQUID loop modulates the effective Josephson coupling energy from island to island, according to 
\begin{equation}\label{eq:Ejay}
E_J(\Phi) = 2 E_J\left|\cos\left(\frac{\pi\Phi}{\Phi_0}\right)\right|\;,
\end{equation}
where $E_J$ is the Josephson energy of a single junction. 
In agreement with previous observations~\cite{Roland_unpublished}, we find $G_V\propto\cos^2(\pi\Phi/\Phi_0)$, thus indicating $G_V\propto E_J^2$ (see Fig.~\ref{IV}(c)). By comparing the data with this relation, we determine the  flux per SQUID for a given external magnetic field.

\begin{figure*}[!htb]
    \begin{center}
        \subfigure[\label{res_logb} Flux-dependent part $\gamma(T)$ of $G_T$]{\includegraphics[width=.4\textwidth]{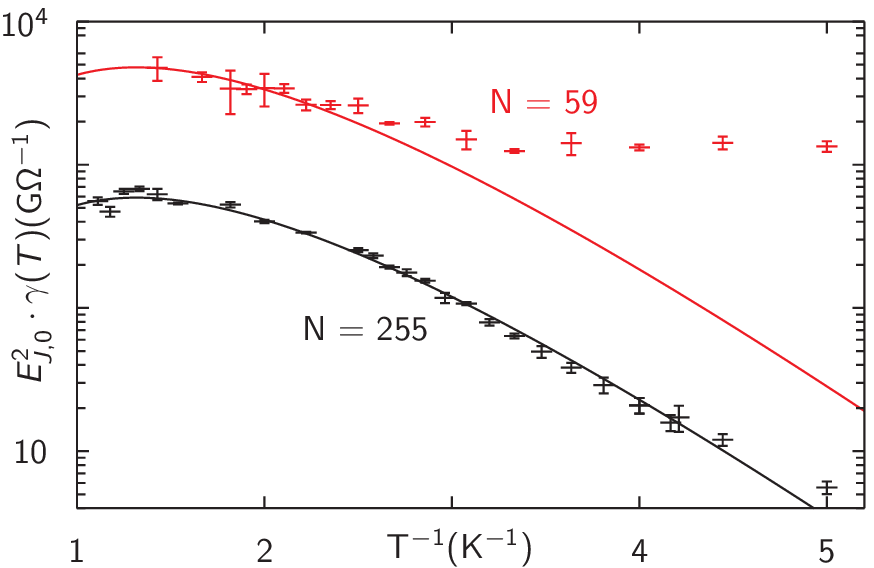}}
	\subfigure[\label{res_loga} Flux-independent part $G_0(T)$ of $G_T$]{\includegraphics[width=.41\textwidth]{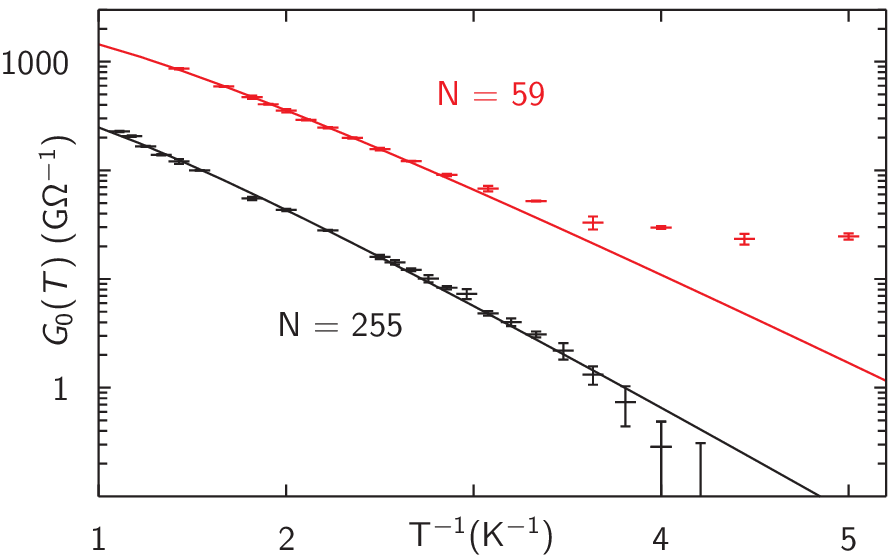}}
    \end{center}
\caption{
Temperature dependence of the zero-bias conductance $G_T$ for both arrays. a) ${E_J}^2$-dependent part ($\gamma(T)$). b) Flux-independent part. The lines represent fits according to Eq.~(\ref{eq_G_T}). The conductance of the array with N=59 islands shows a temperature-independent plateau.
}
\end{figure*}

In the following, we focus on the thermally activated conductance $G_T$ at near-zero bias voltage in the voltage range depicted in the inset of Fig.~\ref{IV}(a). $G_T$ is derived from $I-V$ measurements by employing an odd polynomial function fit  of the 3rd order and extracting the corresponding conductance from the 1st order term. Figure~\ref{IV}(b) shows $G_T$ values measured at different temperatures and magnetic flux. 
At base temperature and a flux close to $\Phi_0/2$, due to the Coulomb blockade of current the conductance is below the current sensitivity threshold of the setup.
At elevated temperatures, a small but measurable, temperature dependent conductance $G_0(T)$ develops at this magnetic flux.
We express the total conductance as a sum of $G_0(T)$ and a flux-dependent part $G_\Phi(T)$,
\begin{equation}\label{eq_separation1}
G_T({\Phi},T) =  G_{\Phi}(T)+ G_0(T)\;,
\end{equation}
and analyze them separately.
The measured data for the conductance $G_{\Phi}$  shows a flux dependence which is well described by a quadratic dependence on the Josephson energy $E_J$ (compare Eq.~(\ref{eq:Ejay})):
\begin{equation}\label{eq_separation}
G_{\Phi}(T) = (2 E_J)^2 \cos^2\left(\frac{\pi\Phi}{\Phi_0}\right) \gamma(T)\;, 
\end{equation}
where the explicit temperature dependence is expressed by $\gamma(T)$. $2 E_J$ is the SQUID Josephson coupling energy at zero magnetic flux.

The least-square fitting of $\gamma(T)$ to Eq.~(\ref{eq_G_T}) yields an activation temperature of $T_0 = 2.66\;\mathrm{K}$ with the temperature dependence of the pre-factor given by $\alpha = 3.5$, see Fig.~\ref{res_logb}. As discussed in the Appendix of this paper, the thermal energy $k_B T_0$ corresponds to the highest on-site energy of the charge carriers in the array $E_{\rm max}$. 
The value $k_B T_0 = E_\mathrm{max} = 229\;\mu\mathrm{eV}$ resulting from the fits is of the same order of magnitude as $\Lambda E_C = 440\;\mu\mathrm{eV}$.
In fact, it is close to the weak disorder estimate of the array charging energy of $\Lambda E_C /2 = 220\;\mu\mathrm{eV}$. 
Thus, the described simple activation mechanism  captures the most essential characteristics of the charge transport in the longer array with $N=255$ islands. 
Above 400~mK, a curve with the same temperature dependence parameters shows reasonable agreement with the data measured for the $N=59$ array.

The least-square fitting of the flux-independent conductance contribution $G_0(T)$ according to Eq.~(\ref{eq_G_T}) yields a temperature dependence pre-factor $\alpha = 1$, see Fig.~\ref{res_loga}. The N=255 array data are best fit with $E_\mathrm{max} = 210\;\mu\mathrm{eV}$. For the $N=59$ array,  $E_\mathrm{max} = 180\;\mu\mathrm{eV}$ is obtained. Interestingly, these values are comparable to the activation energy of the flux-dependent contribution $G_\Phi(T)$. 

We note that the short array with $N=59$ islands shows a temperature-independent plateau conductance in both the flux-dependent and the flux-independent part at low temperatures. This can be due to the influence of picked up voltage noise penetrating further into the interior of the short array, which itself is in terms of electromagnetic screening length only a few $\Lambda$ long. At sufficiently low temperatures, noise-activated conductance may thus dominate.  

It is worth noting that the characteristic energies $E_\mathrm{max}$ obtained from the fits of the flux-independent conductance contribution $G_0(T)$ are rather close to both the superconducting energy gap of thin-film aluminum, $\Delta \approx 200\;\mu\mathrm{eV}$, and the activation energy of the flux-dependent part, $E_\mathrm{max} = 229\;\mu\mathrm{eV}$. One may therefore speculate that this contribution to the conductance is due to thermally generated quasiparticles, or a Cooper pair transport mechanism that is not entirely suppressed, or a combination of both.

\section{Conclusion}

In this paper, we present measurements of the zero-bias conductance of two different arrays of small-capacitance SQUIDs  in dependence of temperature and magnetic flux.   In the analysis, the total conductance $G_T$ is represented by a sum of two parts, a flux-dependent $G_\Phi$  and a flux-independent part $G_0$, where $G_\Phi$ depends quadratically on the Josephson energy $E_J$.  The measured data are consistent with a simple nearest neighbor hopping model having the activation energy $E_\mathrm{max}$ of the order of the charging energy $\Lambda E_C$. This implies that the dominant electrical transport mechanism in the array is the thermal activation of localized charges screened on the characteristic distance of $\Lambda$ islands of the array.

\section{Acknowledgments}

The authors would like to acknowledge stimulating discussions with M. Fistul, D. Haviland, and A. Zorin. This work was supported in part by the EU project SCOPE, the Deutsche Forschungsgemeinschaft (DFG) and the
State of Baden-W{\"u}rttemberg through the DFG Center for Functional Nanostructures (CFN).

\appendix*

\section{Calculation of conductance from the inter-island transition rates}
\label{rederive}

In this Appendix, we rederive and discuss the well-known results for the activational conductance\cite{Gantmakher:book,Shklovskii:book} in the context of transport in a chain of Josephson junctions in the regime of the nearest-neighbor hopping.

Each superconducting island is characterized by the occupation number $f_r\ll1$ of (anti)bosons,
which in equilibrium corresponds to the Boltzmann distribution $f_r^0\approx\exp(-E_r/T)$.
The effective chemical potential is zero due to the presence of the two types of the charge carriers, which can mutually annihilate.
In what follows, we consider the conduction of only one of the species, either bosons or antibosons.

Assume, the system is connected to two reservoirs kept at different chemical potentials.
The resistance of a sufficiently long disordered chain of junctions is dominated by the bulk of the system and is independent
of the nature of the reservoirs and the reservoir-chain couplings. We may assume
that the reservoirs are the islands $r=0$ and $r=N+1$, 
and a non-equilibrium distribution function
\begin{equation}
	f_0=\exp{[-(E_0+2eV)/T]},
	\label{f0}
\end{equation}
is being maintained on the former, while the latter is being kept in equilibrium,
which results in the flow of a stationary current
\begin{equation}
	I=(2e)\left(f_r\Gamma_{r\rightarrow r+1}-f_{r+1}\Gamma_{r+1\rightarrow r}\right),
	\; r=0,\ldots N,
	\label{Current}
\end{equation}
where $\Gamma_{r\rightarrow r+1}$ and $\Gamma_{r+1\rightarrow r}$ are respectively the excitation
transition rates from site $r$ to $r+1$ and from $r+1$ to $r$.

Without loss of generality, we may assume that the current is caused by the diffusion of the charge carriers and
disregard the electric field due to the redistribution of the charges, which otherwise could have been accounted
for by introducing a generalized electrochemical potential in place of the chemical potential on the islands\cite{Shklovskii:book}. Thus, the transition rates in presence of the current coincide with those in equilibrium, and
it holds
\begin{eqnarray}
	e^{-\frac{E_r}{T}} \Gamma_{r\rightarrow r+1} = e^{-\frac{E_{r+1}}{T}} \Gamma_{r+1\rightarrow r}
	\equiv\gamma_{r,r+1}.
	\label{DetailedBalance}
\end{eqnarray}

From Eqs.~(\ref{Current}) and (\ref{DetailedBalance}) we obtain
\begin{eqnarray}
	&&f_r\exp(E_r/T)-f_{r+1}\exp(E_{r+1}/T)=I/(2e\gamma_{r,r+1}),
	\nonumber\\
	&&r=0,\ldots ,N.
	\label{finterm}
\end{eqnarray}

Summing up Eqs.~(\ref{finterm}) from $r=0$ to $r=N$ allows one to express the current
in terms of the non-equilibrium distribution functions in the reservoirs:
\begin{equation}
	I=2e\frac{f_0\exp(E_0/T)-f_{N+1}\exp(E_{N+1/T})}{\sum_{r=0}^N\gamma_{r,r+1}^{-1}}.
	\label{CurrentGeneral}
\end{equation}

In the limit $V\rightarrow0$ we find for the conductance
\begin{equation}
	G=\frac{(2e)^2}{T\sum_{r=0}^N\gamma_{r,r+1}^{-1}},
\end{equation}
which, rather expectedly, corresponds  to the conductance of a series of resistors with
resistances
\begin{equation}
	R_{r,r+1}=\frac{T}{(2e)^2\gamma_{r,r+1}}.
\end{equation}

Taking into account Eq.~(\ref{DetailedBalance}), we obtain $R_{r,r+1}\propto\exp\left[\max(E_r,E_{r+1})/T\right]$,
in agreement with Eq.~(\ref{AMresistor}), which, in the limit $|E_r-E_{r+1}|\gg T$ under consideration,
can be understood as follows.
The rate $\gamma_{sp}$ of spontaneous emission from the higher- to the lower-energy state is temperature independent,
which leads to $\gamma_{r,r+1} = \gamma_{sp}\exp[{\max(E_r,E_{r+1}/T})]$.

\bibliography{literature/lit}

\end{document}